\documentclass[twocolumn,apjl]{openjournal}

\newcommand{\neon}{{$^{22}$Ne}}

\newcommand{\Gaia}{{\it Gaia}}
\newcommand{\be}{\begin{equation}}
\newcommand{\ee}{\end{equation}}

\usepackage{xcolor}
\definecolor{my_color}{HTML}{3a18b1}

\definecolor{new_color}{HTML}{CF0000}

\definecolor{new_black}{HTML}{000000}

\newcommand{\msun}{M$_\odot$}

\usepackage{xcolor}
\usepackage{hyperref}
\usepackage{breakurl}
\usepackage{amsmath}
\usepackage{graphicx}
\usepackage{verbatim}
\usepackage{booktabs}
\usepackage{orcidlink}

\hypersetup{
    colorlinks,
    linkcolor={red!50!black},
    citecolor={blue!50!black},
    urlcolor={blue!80!black}
}

\slugcomment{}

\shorttitle{Enhanced WD Habitability due to \neon\ Distillation}
\shortauthors{Vanderburg et al. 2024}

\begin{document}



\title{Long-lived Habitable Zones around White Dwarfs undergoing Neon-22 Distillation}

\author{Andrew Vanderburg\altaffilmark{1,$\star$}\orcidlink{0000-0001-7246-5438}, Antoine B\'edard\altaffilmark{2}\orcidlink{0000-0002-2384-1326}, Juliette C. Becker\altaffilmark{3}\orcidlink{0000-0002-7733-4522}, Simon Blouin\altaffilmark{4}\orcidlink{0000-0002-9632-1436}}
%

\altaffiltext{1}{Department of Physics and Kavli Institute for Astrophysics and Space Research, Massachusetts Institute of Technology, Cambridge, MA 02139, USA}
\altaffiltext{$\star$}{\url{andrewv@mit.edu}, Sloan Research Fellow}
\altaffiltext{2}{Department of Physics, University of Warwick, CV4 7AL, Coventry, UK}
\altaffiltext{3}{Department of Astronomy,  University of Wisconsin-Madison, 475 N.~Charter St., Madison, WI 53706, USA}
\altaffiltext{4}{Department of Physics and Astronomy, University of Victoria, Victoria, BC V8W 2Y2, Canada}


\begin{abstract}

White dwarf stars have attracted considerable attention in the past 15 years as hosts for potentially habitable planets, but their low luminosity and continuous cooling are major challenges for habitability. Recently, astronomers have found that about 6\% of massive white dwarfs seem to have ``paused'' their cooling for up to $\sim$10\,Gyr. The leading explanation for this cooling delay is the distillation of neutron-rich isotopes such as \neon\ in the white dwarf's interior, which releases a considerable amount of gravitational energy as the star's internal structure rearranges. Here, we consider the impact of \neon\ distillation on the evolution of white dwarf habitable zones. We find that \neon\ distillation in the white dwarf host dramatically increases the time that a planet can continuously reside within the habitable zone (giving more time for life to arise) and that long-lasting habitable zones are located farther from the star (decreasing the impact of tidal forces). These properties may make white dwarfs undergoing \neon\ distillation more promising locations for habitability than white dwarfs undergoing standard cooling.  

\end{abstract}


\section{Introduction}

\begin{figure*}[t!]  
    \centering
    \includegraphics[width=0.98\textwidth]{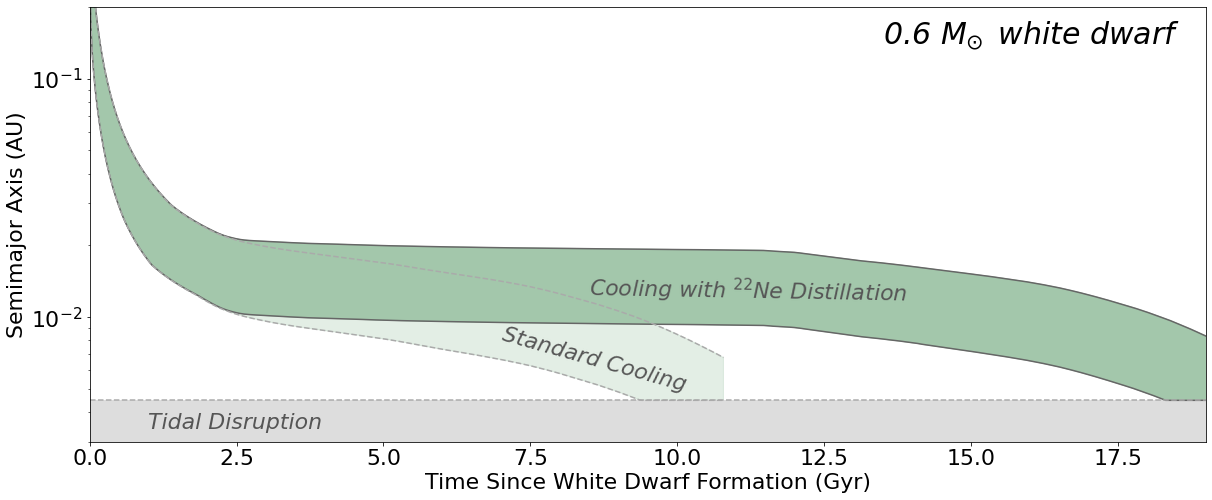}
    \caption{Location of the habitable zone over time for 0.6\,\msun\ white dwarfs cooling with and without \neon\ distillation. We color the region within the habitable zone dark green for white dwarfs with \neon\ distillation and transparent light green for white dwarfs undergoing standard cooling. We show the outlines for white dwarfs with and without \neon\ distillation in solid and dashed lines, respectively. At the bottom of the plot, we show the region of the system where Earth-like planets would experience tidal disruption in solid grey. \neon\ distillation visibly changes the evolution of the habitable zone by introducing a roughly 10\,Gyr pause in its movement inwards. }
    \label{hzcoolingcurve}
\end{figure*}

White dwarfs are the end states of most stars in our galaxy. They are dense, degenerate stellar remnants that form after low to intermediate mass stars exhaust their hydrogen fuel, puff up into a red giant, shed their outer layers, and reveal their hot inert cores. After formation, white dwarfs do not produce energy via nuclear fusion, yet they still shine with the leftover heat from nuclear reactions in their progenitors' cores. As a result, white dwarfs cool over time and eventually (on timescales considerably longer than the age of the universe) will fade into darkness. Typical white dwarfs are about $10^2$ times smaller and roughly $10^2$ to $10^4$ times fainter than the Sun. 

The small size and low luminosity of white dwarfs makes them interesting locations to search for orbiting planets. Compared to main sequence stars, it is considerably easier to detect the transits of planets orbiting white dwarfs \citep{DiStefano2010ApJ, Faedi2011MNRAS, agol2011, Fulton2014ApJ}, transmission signals from the atmospheres of those planets \citep{Loeb2013MNRAS, Lin2014ApJLpollution, Kaltenegger2020ApJL}, and thermal emission from either resolved or unresolved companions \citep{Lin2014ApJLasteroid, Limbach2022MNRAS, Poulsen2024AJ}. Moreover, because white dwarfs are faint, their habitable zones are very close ($\sim$10$^{-2}$\,AU), making it possible to observe many planetary orbits on short timescales. The prospect of detecting and studying the atmospheres of rocky planets in the habitable zones of white dwarfs is highly enticing, because should such planets exist, they would offer a shortcut towards searching for biosignatures in the atmosphere of an Earth-sized planet \citep{agol2011, Loeb2013MNRAS, Kozakis2020ApJL, Kaltenegger2020ApJL, Lin2022ApJL, Doshi2022MNRAS, Whyte2024arXiv}. 

As a result, in recent years, the study of planetary systems around white dwarfs has gained popularity \citep{Veras2021orel}. Searches for exoplanets orbiting the main-sequence progenitors of white dwarfs have found that planets are a nearly ubiquitous outcome of star formation \citep[e.g.,][]{Mayor2011arXiv, Ghezzi2018ApJ, Dattilo2023AJ}. Although the planets orbiting close to their main sequence hosts will be destroyed during the hosts' post-main-sequence evolution, planets orbiting more than a few AU from their stars are expected to survive \citep{Mustill2012ApJ, Nordhaus2013MNRAS} and continue orbiting the white dwarf \citep[e.g.,][]{luhman2011, Blackman2021Natur, Mullally2024ApJL}. Interestingly, there is also considerable observational evidence that the surviving planetary material can journey inwards towards the white dwarf, ranging from asteroids and comets \citep[e.g.,][]{jura2003, becklin2005, Zuckerman2010ApJ, Kilic2006ApJ, Farihi2013Sci} to minor planets \citep{Dufour2012ApJ, wd1145, Manser2019Sci} to ice or gas giant planets \citep{gansicke2019, wd1856}. One giant planet has been found transiting the white dwarf WD\,1856+534 at an orbital separation of about 0.02\,AU, just outside the white dwarf's habitable zone \citep{wd1856}.  

While we now know that white dwarfs do indeed host planets and that at least some of them can survive migration towards the habitable zone, there remain considerable challenges to the habitability of closely orbiting planets. In order for planets surviving the red giant branch at orbital distance of a few AU to migrate into the habitable zone, they must dissipate much of their orbital energy without being destroyed either inside the stellar envelope \citep{passy} or via tidal disruption \citep{Veras2020MNRAS}. Meanwhile, in order for these surviving planets to retain water, they must avoid dessication by the extreme ultraviolet radiation from the young white dwarf \citep{becker2024}. Once the planets do arrive close to the white dwarf, they will find an inward-moving habitable zone due to the white dwarf's cooling; only after the white dwarf has cooled below $\sim$10,000\,K and the habitable zone has moved to around $10^{-2}$\,AU does it evolve slowly enough for a stationary planet to remain habitable for more than 3\,Gyr \citep{agol2011}. And finally, if a planet is in a long-lived habitable zone close to the white dwarf, it will experience extreme tidal forces that can quickly dissipate any eccentricity forced by additional planets in the system, potentially heating the planet until it is no longer hospitable to life \citep{barnesheller, Becker2023ApJL}.  

\setcounter{footnote}{4}

In this paper, we identify a physical process taking place in some white dwarfs that may offer the opportunity for enhanced habitability compared to normal white dwarf evolution: cooling delays caused by the distillation of neutron-rich isotopes, most notably \neon. Recently, an overdensity of massive white dwarfs on the HR diagram called the ``Q-branch'' was identified in \Gaia\ data \citep{GaiaCollaboration2018A&A} in a region associated with delays to white dwarf cooling due to crystallization \citep{Tremblay2019Natur}. Analysis of the HR diagram and the Q-branch stars' kinematics by \citet{Cheng2019ApJ} indicated that about 6\% of ultramassive ($\gtrsim 1$\,\msun) white dwarfs effectively pause their cooling for at least 8\,Gyr. The leading physical explanation for this phenomenon is the formation and movement of \neon-poor crystals in the interiors of white dwarfs \citep{Blouin2021ApJL, Bedard2024Natur}. The crystals are buoyant, so their formation leads to a rearrangement of the interior structure of the star, releasing considerable gravitational potential energy. This process should not take place in all white dwarfs --- only those with an enhanced abundance of \neon\ (greater than about 2.5\% by mass)\footnote{\citet{Blouin2021ApJL} suggested that distillation may also take place in white dwarfs with a standard \neon\ content. In this case, distillation would only start after $\sim$60\% of the core has solidified normally and would then occur in a thin shell around the solid core, leading to a shorter cooling delay of $\sim$1\,Gyr \citep[see also][]{segretain1996}. In the present paper, we ignore this ``shell distillation'' and focus exclusively on the ``core distillation'' experienced by \neon-enhanced white dwarfs.} either due to a high primordial $\alpha$ element abundance or white dwarf mergers \citep{Bauer2020ApJ, Shen2023ApJL, Salaris2024} --- explaining why only about 6\% of ultramassive white dwarfs experience a delayed cooling. We expect that most \neon-distilling stars hosting habitable planets will have achieved \neon-enhancement from their primordial abundances, since the explosions resulting from white dwarf merger events presumably are harmful to habitability in the system. We note, however, that debris from white dwarf mergers could form second generation planets \citep[e.g.,][]{Wolszczan1992Natur, Livio2005ApJL}, so habitable planets around merger remnants are not completely out of the question.   

Our paper is organized as follows. In Section\,\ref{models}, we describe the generation of white dwarf cooling models incorporating \neon\ distillation resulting from a high primordial metallicity. Section\,\ref{hzlifetimes} describes our calculation of the location and duration of the continuous habitable zone around white dwarfs for both the \neon-distillation and standard cooling scenarios. Section\,\ref{water} discusses how \neon\ distillation affects a planet's prospects for water retention. Finally, in Section\,\ref{discussion}, we discuss the implications of this work for habitability around white dwarfs and conclude. 

\begin{deluxetable*}{cccccccc}
\centering
\tabletypesize{\small}
\tablewidth{0pt}
\tablecaption{Continuous Habitable Zone (CHZ) Maximum Duration and Location}
\tablehead{\colhead{White Dwarf} & \colhead{Progenitor} & \colhead{Model} & \colhead{$T_{\rm eff}$ at} & \colhead{Standard CHZ} & \colhead{Standard CHZ} & \colhead{\neon\ CHZ} & \colhead{\neon\ CHZ} \\\colhead{Mass (\msun)} & \colhead{Mass (\msun)} & \colhead{Type} &\colhead{Pause (K)} & \colhead{Duration (Gyr)} & \colhead{Location (AU)} & \colhead{Duration (Gyr)} & \colhead{Location (AU)}}
\startdata
0.6 & 2.01 & Single Star & 5,800 & 6.67 & 0.006--0.0183 & 15.56 & 0.006--0.0203 \\
0.8 & 3.48 & Single Star & 8,600 & 6.52 & 0.0068--0.0183 & 12.91 & 0.0067--0.0305 \\
1.0 & 5.76 & Single Star & 12,900 & 4.69 & 0.0075--0.0145 & 7.39 & 0.0074--0.0156, 0.0223--0.0533 \\

1.0 & N/A & Merger & 10,500 & 6.39 & 0.0072--0.0319 & 16.45 & 0.0071--0.038 \\
1.1 & N/A & Merger & 13,000 & 4.77 & 0.0091--0.0376 & 13.91 & 0.0087--0.05 \\
1.2 & N/A & Merger & 17,300 & 3.72 & 0.0208--0.044 & 10.38 & 0.0296--0.071 

\enddata
\tablecomments{Note: We define the CHZ as locations where stationary orbiting planets would remain in the habitable zone for at least 3\,Gyr. To estimate the progenitor masses for white dwarfs that formed via single-star evolution, we use the Initial/Final Mass Relation (IFMR) of \citet{El-Badry2018ApJL}. We report the stellar effective temperature at the onset of \neon\ distillation in the column labeled ``$T_{\rm eff}$ at Pause.''  \label{hzlocations}}
\end{deluxetable*}

\section{White Dwarf Cooling Models}\label{models}

We generated white dwarf cooling models including \neon\ distillation using the STELUM code \citep{Bedard2022}. We relied on the same setup described in \citet{Bedard2024Natur}, except that we considered different masses and compositions. \citet{Bedard2024Natur} focused on ultramassive white dwarfs with atypical compositions owing to a merger history, in particular thin hydrogen and helium layers of masses $M_{\mathrm{H}}/M_* = 10^{-10}$ and $M_{\mathrm{He}}/M_* = 10^{-6}$, respectively. For this work, we extended the model grid to lower stellar masses and considered compositions resulting from (high-metallicity) single-star evolution. We considered 0.6, 0.8, and 1.0\,\msun\ white dwarfs, for which we assumed ($\log M_{\mathrm{H}}/M_*$, $\log M_{\mathrm{He}}/M_*$) = ($-4.0$, $-2.0$), ($-5.0$, $-2.5$), and ($-6.0$, $-3.0$), respectively \citep{Romero2013}. For the initial core composition, we adopted the oxygen abundance profiles of \citet{Bauer2023} and a uniform \neon\ abundance of 3\% (by mass). For comparison, we also computed an analogous set of models with a lower abundance of 1.5\% (as expected from solar-metallicity progenitors) and thus without core distillation. We refer the reader to \citet{Bedard2024Natur} for more details on the model calculations.

We find that our 0.6, 0.8, and 1.0\,\msun\ models experience cooling delays of about 10, 9, and 6\,Gyr at ($T_{\rm eff}$/K, $\log L/L_{\odot}$) $\simeq$ (5500, $-3.9$), (8100, $-3.4$), and (12,000, $-2.9$), respectively. For comparison, the 1.0\,\msun\ merger-type model of \citet{Bedard2024Natur} experiences a cooling delay of about 13\,Gyr at ($T_{\rm eff}$/K, $\log L/L_{\odot}$) $\simeq$ (10,000, $-3.2$). The difference is mainly due to the thinner helium layer, which causes distillation to occur at a lower luminosity and thus increases the time required to spend the extra energy \citep{Bedard2024Natur}.

\section{Continuous Habitable Zone Location and Duration}\label{hzlifetimes}

Using our white dwarf cooling models with and without \neon\ distillation, we studied how this process affects the evolution of white dwarf habitable zones. The dominant factor that determines whether a planet is in the habitable zone (that is, whether it can sustain liquid water on its surface) is the incident flux received by the planet, which is determined by the luminosity of the star. However, changing the spectrum of the incident flux can also affect habitable zone boundaries, even when the overall flux level is unchanged. Therefore, the temperature of the star is often included in habitable zone models as a second order correction.  

We estimated the outer and inner boundaries of the habitable zone based on the work of \citet{Kopparapu2013ApJ} and \citet{Zhan2024arXiv}, respectively. For the outer boundary, we use the polynomial fitting functions provided by  \citet{Kopparapu2013ApJ} that approximate the minimum incident flux received by planets capable of sustaining a greenhouse effect large enough for liquid water to exist on a planet's surface (the so-called ``conservative'' habitable zone boundary). To determine the inner edge of the habitable zone, we use the results of global climate models (GCMs) from \citet{Zhan2024arXiv}. This team recently discovered that terrestrial planets around white dwarfs can experience a novel atmospheric circulation pattern, dubbed ``bat rotation,'' that permits liquid water to exist at higher incident fluxes than around main sequence stars (see also \citealt{Shields2024arXiv}). Following \citet{Zhan2024arXiv}, we define a piecewise linear function based on the input parameters for the GCM simulation at each stellar effective temperature with the highest incident flux to avoid a runaway greenhouse effect (see Figure\,14 of \citealt{Zhan2024arXiv}). 


We calculated the habitable zone boundaries at each time step for the white dwarf cooling tracks described in Section\,\ref{models} as well as cooling tracks for merger remnants from \citet{Bedard2024Natur}. The fitting functions provided by \citet{Kopparapu2013ApJ} are calibrated for stars with temperatures between 2,600 and 7,200\,K. At early times in our white dwarf cooling tracks, the effective temperature is significantly higher than 7,200\,K, so we assumed the same flux boundaries as calculated by \citet{Kopparapu2013ApJ} at 7,200\,K, effectively ignoring the second-order correction due to the different spectral shape of flux from higher temperature white dwarfs. At late times, the inner edge of the habitable zone can move inside the tidal disruption radius (also known as the Roche limit) for Earth-like planets. We calculate this minimum semimajor axis for planetary survival following \citet{Rappaport2013ApJL} and substitute this value for the inner edge of the habitable zone when the flux-based habitable zone definition falls below this distance. 

We find that the cooling delay from \neon\ distillation has a dramatic impact on the evolution of white dwarf habitable zones. An example of the effect of \neon\ distillation on the habitable zone surrounding a 0.6\,\msun\ white dwarf is shown in Figure\,\ref{hzcoolingcurve}. In standard white dwarf cooling, the boundaries of the habitable zone continuously move inward; planets must be closer to the star to maintain liquid water on the surface as the star's energy output decreases. But while \neon\ distillation is taking place during the white dwarf's cooling, the star's luminosity and temperature remain effectively constant, and the habitable zone's evolution pauses for about 10\,Gyr. 

We performed similar calculations for white dwarfs ranging in mass from 0.6 to 1.0\,\msun\ (for high-metallicity single-star compositions) and from 1.0 to 1.2\,\msun\ (for merger remnant compositions) both with and without \neon\ distillation. For each of these calculations, we measured the longest time that any particular location remained within the habitable zone. We found that \neon\ distillation consistently increases this duration by a factor of about 2--3 compared to standard white dwarf cooling. We also determined the location of the ``continuous habitable zone'' --- that is, the regions in the system that remain in the habitable zone for longer than 3\,Gyr. We find that the range in semimajor axis where orbiting planets will experience at least 3\,Gyr in the habitable zone is considerably increased when including \neon\ distillation in the cooling tracks. In particular, while the inner edge of the continuous habitable zone is similar to the standard cooling case, the outer edge extends farther away from the white dwarf than the $10^{-2}$\,AU typical distance identified by \citet{agol2011}. In some cases for our \neon\ cooling models, we notice there are two distinct regions of the system that spend longer than 3\,Gyr in the habitable zone: one region that corresponds to the \neon\ distillation cooling pause, and a second region occurring after the \neon\ distillation analogous to the continuous habitable zone in standard white dwarf cooling. We summarize the results from these calculations in Table\,\ref{hzlocations}. 



\section{Water retention }\label{water}

In addition to maintaining appropriate surface temperatures, another important factor in assessing habitability prospects for planets orbiting white dwarfs is the feasibility of retaining their surface water. Planets migrating inwards from larger orbital distances --- where they originally resided during the host star's main sequence phase --- are particularly vulnerable to water loss. This can occur through tidal heating, which can drive water evaporation into the planetary atmosphere or Jeans escape at high temperatures, and photoevaporation driven by XUV radiation emitted by the white dwarf. 
In this work, we follow the methodology of \citet{becker2024}, applying the same coupled framework for orbital evolution, tidal heating, and ocean mass loss. The primary difference here compared to that previous work is our adoption of updated white dwarf cooling models for three white dwarf masses ($0.6$, $0.8$, and $1.0\,M_{\odot}$), all of which incorporate the effects of \neon\ distillation (see Section\,\ref{models}). 

\begin{figure}[t!]  
    \centering
    \includegraphics[width=0.45\textwidth]{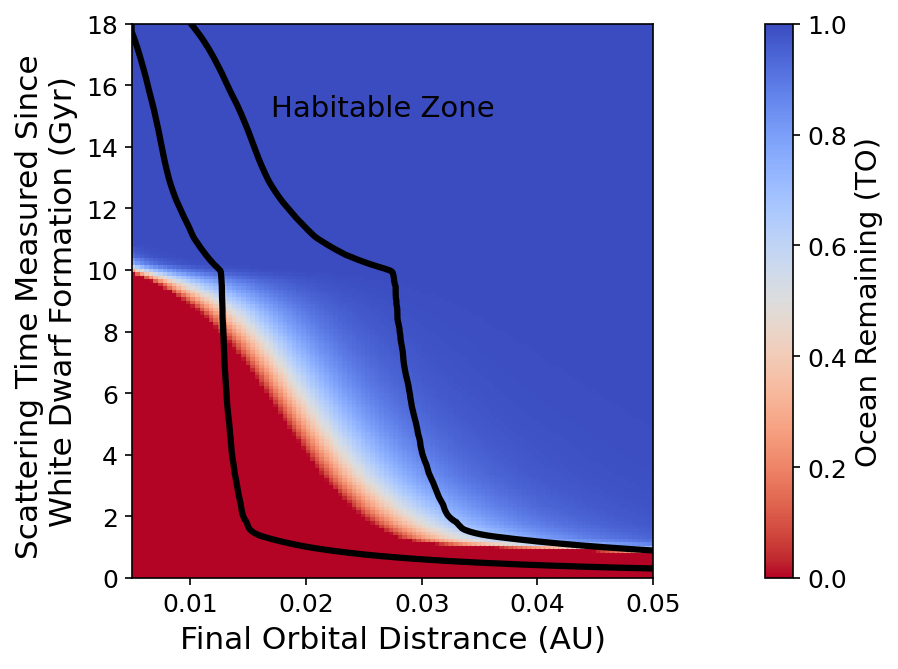}
    \caption{Ocean retention (in terrestrial oceans) for a planet with 1 T.O. of initial surface water orbiting a 0.8\,M$_{\odot}$ white dwarf as a function of its final orbital distance $a_f$ and the planet's time of arrival at its observed orbit (measured since the white dwarf's formation). Red shading represents desiccation of the planetary surface, while blue shading represents substantial retention or surface water. The black lines denote the boundaries of the habitable zone.}
    \label{fig:ocean}
\end{figure}

In brief, to model water retention, we begin with a planet scattered onto a highly eccentric orbit around a white dwarf. The planet’s subsequent orbital evolution is computed using standard tidal evolution equations \citep[e.g.,][]{Hut1981}, which model how the tidal dissipation of orbital energy in the planet’s interior causes its orbital eccentricity $e$ and semi-major axis $a$ to evolve. As the orbit circularizes and shrinks, the intense tidal heating associated with high eccentricities can heat the planet’s surface, potentially evaporating some or all of its surface water. 
During this process, we simultaneously model the planet’s ocean mass loss due to both the heat input from tidal heating and subsequent photoevaporation due to the white dwarf’s XUV radiation. We estimate the XUV ($\lambda < 1200$\,\AA) flux fraction of the white dwarf using pure-hydrogen atmosphere models \citep{Tremblay2011, Blouin2019} interpolated at the effective temperatures and surface gravities of our cooling tracks. 
Using these models, we compute the water retained for a planet orbiting a white dwarf as a function of the planet's final semi-major axis and the time at which it arrives in that orbit. 
We assume a planet with Earth-like mass and radius ($1\,M_{\oplus}$, $1\,R_{\oplus}$) and one terrestrial ocean's  (T.O.) worth of surface water initially placed at a semi-major axis of 5\,AU, and vary its $e$ (and subsequently its initial pericenter distance). 
The result of this parameter sweep is shown in Figure\,\ref{fig:ocean}, and overlaid is the location of the habitable zone as computed in Section\,\ref{hzlifetimes} applied to our white dwarf luminosity models.

We also produce similar parameter sweeps using largely identical initial planetary and orbital parameters for a variety of initial conditions, including three different white dwarf masses ($0.6$, $0.8$, and $1.0\,M_{\odot}$), and two different initial ocean contents (1 and 10 T.O., with assumed planetary tidal quality factor $Q_p=10$ and 100, respectively). In Table\,\ref{tab:ocean}, we present the maximum surface water remaining anywhere in the habitable zone at 3\,Gyr after the white dwarf formed.
These results indicate that although \neon\ distillation may improve the positioning of the habitable zone around a 1.0\,M$_{\odot}$ white dwarf, it remains unlikely that planets residing within the habitable zone will retain surface water. In contrast, for 0.6--0.8\,M$_{\odot}$ white dwarfs, substantial water retention is still possible inside the habitable zone, even for lower initial amounts of surface water. 

\begin{table}[h!]
\centering
\caption{Surface Water Retention for Host White Dwarfs of Various Masses. }
\begin{tabular}{ccc}
\hline
\textbf{White Dwarf Mass} & \multicolumn{2}{c}{\textbf{Ocean Retained by 3\,Gyr}} \\ 
\hline
{M$_\odot$} & { 1 T.O.} & {10 T.O.} \\ \hline
0.6    & 0.99    & 9.99    \\
0.8   & 0.80     & 9.80     \\
1.0       & 0.00      & 0.00       \\
\hline
\end{tabular}
\tablecomments{Results show the maximum amount of water present anywhere in the habitable zone after 3\,Gyr for three different white dwarf masses ($0.6$, $0.8$, and $1.0\,M_{\odot}$). Initial ocean masses are in units of Earth T.O. (Terrestrial Oceans). Planetary tidal $Q_p= 10$ for the 1 T.O. case and $Q_p= 100$ for the 10 T.O. case. Note all of these cases assume compositions from high-metallicity single star evolution.}
\label{tab:ocean}
\end{table}

\section{Discussion and Conclusions}
\label{discussion}

The delayed cooling caused by distillation in \neon-enriched white dwarfs dramatically alters the evolution of white dwarf habitable zones. Remarkably, several of the changes to the habitable zone's evolution work to ease potential challenges to life in white dwarf systems. In particular, \neon\ distillation provides three key advantages to habitability compared to standard white dwarf cooling:  
\begin{enumerate}
    \item \textit{Longer continuous habitable zone lifetimes:} We find that \neon\ distillation tends to increase the maximum amount of time that any location in the system remains in the habitable zone by about a factor of 2--3, and up to about 10\,Gyr. The longer habitable zone lifetime both allows more time for a planet to migrate into the habitable zone, as well as once the planet is in place, more time for life to develop. Moreover, the region of the system that spends at least 3\,Gyr in the habitable zone is considerably expanded, making it more likely that if a planet were to migrate close to the white dwarf, it would spend long enough in the habitable zone for life to arise.  
    \item \textit{Lower importance of tides:} We also find that \neon\ distillation increases the semimajor axis of the outer edge of the continuous habitable zone by over 50\%. At larger distances from the host star, tidal effects from the star grow rapidly less important; tidal heating scales with orbital radius to the $-7.5$ power \citep{Goldreich1966}. It is therefore considerably easier for planets orbiting farther from the white dwarf to avoid the ``tidal greenhouse'' phenomenon identified by \citet{barnesheller}. These authors pointed out that in the $\sim$10$^{-2}$\,AU continuous habitable zone identified by \citet{agol2011}, even a small forced eccentricity of $e\sim10^{-5}$ could generate enough tidal heating to render the planet uninhabitable. Increasing the semimajor axis by 50\% decreases the tidal heating by a factor of 20 and makes it easier to avoid tidal desiccation. 
    \item \textit{Greater stability of incident flux:} Finally, the pause in cooling during \neon\ distillation stabilizes the climate of the planet. Even though the changes to the incident flux on a planet in the continuous habitable zone around a normally cooling white dwarf would be slow compared to the evolutionary timescale of life on Earth, the stability of the climate likely simplifies the conditions needed for life to arise and thrive for many billions of years. 

\end{enumerate}
In addition to these three impacts that are likely advantageous for habitability in white dwarf systems, one other effect of \neon\ distillation cooling delays might be harmful: the newly-expanded continuous habitable zone exists earlier in the white dwarf's lifetime, when the effective temperature of the remnant is considerably higher than the Sun (see Table\,\ref{hzlocations}). At temperatures higher than about 10,000\,K, it is difficult to avoid considerable loss of water from the planet due to photoevaporation \citep{barnesheller, becker2024}. Our modeling indicates that the pause in cooling for $\gtrsim$ 1\,\msun\ white dwarfs will take place at high enough stellar temperatures that water retention will be difficult. However, this impact is less problematic for lower-mass white dwarfs, which undergo \neon\ distillation after the white dwarf has cooled to lower temperatures. 

An important open question in the field of white dwarf research is how did the stars undergoing \neon\ distillation end up with such a high concentration of \neon\ in their interiors? Typical solar-neighborhood compositions for stars are expected to yield a \neon\ mass fraction of about 1.5\%, which is not enough to trigger core distillation. Instead, a sufficient \neon\ enhancement likely requires either a high primordial abundance of $\alpha$ elements or further nuclear processing through a sub-Chandrasekhar white dwarf merger \citep{Bauer2020ApJ, Blouin2021ApJL, Shen2023ApJL}. Evidence is emerging that many of the high-mass members of the Q-branch are indeed merger remnants \citep[e.g.,][]{Hollands2020NatAs, Fleury2022MNRAS, Kilic2024ApJ}. Presumably the explosive merger of two closely orbiting stars is harmful to habitability on existing planetary objects in the system, but perhaps the debris from the explosion may trigger the formation of second generation habitable planets \citep{Wolszczan1992Natur, Livio2005ApJL}. Meanwhile, the fact that the Q-branch is likely populated with merger remnants does not preclude \neon\ distillation in white dwarfs with primordial $\alpha$ element enhancements. In fact, evidence for the latter scenario was recently provided by \citet{Salaris2024}, who showed that distillation is needed to explain the luminosity function of white dwarfs in the high-metallicity ($Z \simeq 0.03$) open cluster NGC 6791. In cases like these, \neon\ distillation may enhance the habitability of more traditional, first-generation planets around white dwarfs.



Finally, this work motivates future calculations to understand the HZ boundaries at high temperatures. A limitation of our calculation is that we were unable to account for how the different spectrum of incident radiation on Earth-like planets from host stars hotter than 7200\,K affects habitable zone boundaries. The subject of habitable zones around hot stars has received relatively little attention because hot main sequence stars have relatively short lifetimes and the duration of the continuous habitable zone was believed to be too short for life to arise \citep{Kasting1993Icar, Kopparapu2013ApJ}. This has also been true of hot white dwarfs; during standard white dwarf cooling, the habitable zone only evolves slowly enough for long-lived habitable zones to exist once the white dwarfs have cooled substantially \citep{agol2011}. Now that we have showed that \neon\ distillation can freeze the rapid evolution of white dwarf habitable zones for up to 10\,Gyr, and that planets can retain water around white dwarfs above 7200\,K, there is a clear situation in which planets could remain in a long-lived habitable zone around such hot stars. Ultimately, if \neon\ distilling white dwarfs are found to indeed host habitable planets, they may expand our understanding of where life can arise in the universe.

\acknowledgments
This work is an outcome of the ideas and conversations shared at the 2024 European Workshop on White Dwarfs (EuroWD24); we thank the conference organizers for creating a compelling scientific program that enabled new ideas to form. We thank Tad Komacek and Mary Anne Limbach for helpful conversations. This research has made use of NASA's Astrophysics Data System. AB is a Postdoctoral Fellow of the Natural Sciences and Engineering Research Council (NSERC) of Canada and also acknowledges support from the European Research Council (ERC) under the European Union’s Horizon 2020 research and innovation program (grant agreement no. 101002408).

Facilities: \facility{ADS}\\

Software: \texttt{matplotlib} \citep{plt},
          \texttt{numpy} \citep{np},
          \texttt{pandas} \citep{mckinney-proc-scipy-2010, reback2020pandas}

\bibliographystyle{apj}
\bibliography{refs}

\begin{thebibliography}{}
\expandafter\ifx\csname natexlab\endcsname\relax\def\natexlab#1{#1}\fi

\bibitem[{{Agol}(2011)}]{agol2011}
{Agol}, E. 2011, \apjl, 731, L31

\bibitem[{{Barnes} \& {Heller}(2013)}]{barnesheller}
{Barnes}, R., \& {Heller}, R. 2013, Astrobiology, 13, 279

\bibitem[{{Bauer}(2023)}]{Bauer2023}
{Bauer}, E.~B. 2023, \apj, 950, 115

\bibitem[{{Bauer} {et~al.}(2020){Bauer}, {Schwab}, {Bildsten}, \& {Cheng}}]{Bauer2020ApJ}
{Bauer}, E.~B., {Schwab}, J., {Bildsten}, L., \& {Cheng}, S. 2020, \apj, 902, 93

\bibitem[{{Becker} {et~al.}(2023){Becker}, {Seligman}, {Adams}, \& {Styczinski}}]{Becker2023ApJL}
{Becker}, J., {Seligman}, D.~Z., {Adams}, F.~C., \& {Styczinski}, M.~J. 2023, \apjl, 945, L24

\bibitem[{{Becker} {et~al.}(2024){Becker}, {Vanderburg}, \& {Livesey}}]{becker2024}
{Becker}, J., {Vanderburg}, A., \& {Livesey}, J. 2024, arXiv e-prints, arXiv:2412.12056

\bibitem[{{Becklin} {et~al.}(2005){Becklin}, {Farihi}, {Jura}, {Song}, {Weinberger}, \& {Zuckerman}}]{becklin2005}
{Becklin}, E.~E., {Farihi}, J., {Jura}, M., {et~al.} 2005, \apjl, 632, L119

\bibitem[{{B{\'e}dard} {et~al.}(2024){B{\'e}dard}, {Blouin}, \& {Cheng}}]{Bedard2024Natur}
{B{\'e}dard}, A., {Blouin}, S., \& {Cheng}, S. 2024, \nat, 627, 286

\bibitem[{{B{\'e}dard} {et~al.}(2022){B{\'e}dard}, {Brassard}, {Bergeron}, \& {Blouin}}]{Bedard2022}
{B{\'e}dard}, A., {Brassard}, P., {Bergeron}, P., \& {Blouin}, S. 2022, \apj, 927, 128

\bibitem[{{Blackman} {et~al.}(2021){Blackman}, {Beaulieu}, {Bennett}, {Danielski}, {Alard}, {Cole}, {Vandorou}, {Ranc}, {Terry}, {Bhattacharya}, {Bond}, {Bachelet}, {Veras}, {Koshimoto}, {Batista}, \& {Marquette}}]{Blackman2021Natur}
{Blackman}, J.~W., {Beaulieu}, J.~P., {Bennett}, D.~P., {et~al.} 2021, \nat, 598, 272

\bibitem[{{Blouin} {et~al.}(2021){Blouin}, {Daligault}, \& {Saumon}}]{Blouin2021ApJL}
{Blouin}, S., {Daligault}, J., \& {Saumon}, D. 2021, \apjl, 911, L5

\bibitem[{{Blouin} {et~al.}(2019){Blouin}, {Dufour}, {Thibeault}, \& {Allard}}]{Blouin2019}
{Blouin}, S., {Dufour}, P., {Thibeault}, C., \& {Allard}, N.~F. 2019, \apj, 878, 63

\bibitem[{{Cheng} {et~al.}(2019){Cheng}, {Cummings}, \& {M{\'e}nard}}]{Cheng2019ApJ}
{Cheng}, S., {Cummings}, J.~D., \& {M{\'e}nard}, B. 2019, \apj, 886, 100

\bibitem[{{Dattilo} {et~al.}(2023){Dattilo}, {Batalha}, \& {Bryson}}]{Dattilo2023AJ}
{Dattilo}, A., {Batalha}, N.~M., \& {Bryson}, S. 2023, \aj, 166, 122

\bibitem[{{Di Stefano} {et~al.}(2010){Di Stefano}, {Howell}, \& {Kawaler}}]{DiStefano2010ApJ}
{Di Stefano}, R., {Howell}, S.~B., \& {Kawaler}, S.~D. 2010, \apj, 712, 142

\bibitem[{{Doshi} {et~al.}(2022){Doshi}, {Cowan}, \& {Huang}}]{Doshi2022MNRAS}
{Doshi}, D., {Cowan}, N.~B., \& {Huang}, Y. 2022, \mnras, 515, 1982

\bibitem[{{Dufour} {et~al.}(2012){Dufour}, {Kilic}, {Fontaine}, {Bergeron}, {Melis}, \& {Bochanski}}]{Dufour2012ApJ}
{Dufour}, P., {Kilic}, M., {Fontaine}, G., {et~al.} 2012, \apj, 749, 6

\bibitem[{{El-Badry} {et~al.}(2018){El-Badry}, {Rix}, \& {Weisz}}]{El-Badry2018ApJL}
{El-Badry}, K., {Rix}, H.-W., \& {Weisz}, D.~R. 2018, \apjl, 860, L17

\bibitem[{{Faedi} {et~al.}(2011){Faedi}, {West}, {Burleigh}, {Goad}, \& {Hebb}}]{Faedi2011MNRAS}
{Faedi}, F., {West}, R.~G., {Burleigh}, M.~R., {Goad}, M.~R., \& {Hebb}, L. 2011, \mnras, 410, 899

\bibitem[{{Farihi} {et~al.}(2013){Farihi}, {G{\"a}nsicke}, \& {Koester}}]{Farihi2013Sci}
{Farihi}, J., {G{\"a}nsicke}, B.~T., \& {Koester}, D. 2013, Science, 342, 218

\bibitem[{{Fleury} {et~al.}(2022){Fleury}, {Caiazzo}, \& {Heyl}}]{Fleury2022MNRAS}
{Fleury}, L., {Caiazzo}, I., \& {Heyl}, J. 2022, \mnras, 511, 5984

\bibitem[{{Fulton} {et~al.}(2014){Fulton}, {Tonry}, {Flewelling}, {Burgett}, {Chambers}, {Hodapp}, {Huber}, {Kaiser}, {Wainscoat}, \& {Waters}}]{Fulton2014ApJ}
{Fulton}, B.~J., {Tonry}, J.~L., {Flewelling}, H., {et~al.} 2014, \apj, 796, 114

\bibitem[{{Gaia Collaboration} {et~al.}(2018){Gaia Collaboration}, {Babusiaux}, {van Leeuwen}, {Barstow}, {Jordi}, {Vallenari}, {Bossini}, {Bressan}, {Cantat-Gaudin}, {van Leeuwen}, \& et~al.}]{GaiaCollaboration2018A&A}
{Gaia Collaboration}, {Babusiaux}, C., {van Leeuwen}, F., {et~al.} 2018, \aap, 616, A10

\bibitem[{{G{\"a}nsicke} {et~al.}(2019){G{\"a}nsicke}, {Schreiber}, {Toloza}, {Fusillo}, {Koester}, \& {Manser}}]{gansicke2019}
{G{\"a}nsicke}, B.~T., {Schreiber}, M.~R., {Toloza}, O., {et~al.} 2019, \nat, 576, 61

\bibitem[{{Ghezzi} {et~al.}(2018){Ghezzi}, {Montet}, \& {Johnson}}]{Ghezzi2018ApJ}
{Ghezzi}, L., {Montet}, B.~T., \& {Johnson}, J.~A. 2018, \apj, 860, 109

\bibitem[{{Goldreich} \& {Soter}(1966)}]{Goldreich1966}
{Goldreich}, P., \& {Soter}, S. 1966, \textit{Icarus}, 5, 375

\bibitem[{{Hollands} {et~al.}(2020){Hollands}, {Tremblay}, {G{\"a}nsicke}, {Camisassa}, {Koester}, {Aungwerojwit}, {Chote}, {C{\'o}rsico}, {Dhillon}, {Gentile-Fusillo}, {Hoskin}, {Izquierdo}, {Marsh}, \& {Steeghs}}]{Hollands2020NatAs}
{Hollands}, M.~A., {Tremblay}, P.~E., {G{\"a}nsicke}, B.~T., {et~al.} 2020, Nature Astronomy, 4, 663

\bibitem[{{Hunter}(2007)}]{plt}
{Hunter}, J.~D. 2007, Computing in Science and Engineering, 9, 90

\bibitem[{{Hut}(1981)}]{Hut1981}
{Hut}, P. 1981, \aap, 99, 126

\bibitem[{{Jura}(2003)}]{jura2003}
{Jura}, M. 2003, \apjl, 584, L91

\bibitem[{{Kaltenegger} {et~al.}(2020){Kaltenegger}, {MacDonald}, {Kozakis}, {Lewis}, {Mamajek}, {McDowell}, \& {Vanderburg}}]{Kaltenegger2020ApJL}
{Kaltenegger}, L., {MacDonald}, R.~J., {Kozakis}, T., {et~al.} 2020, \apjl, 901, L1

\bibitem[{{Kasting} {et~al.}(1993){Kasting}, {Whitmire}, \& {Reynolds}}]{Kasting1993Icar}
{Kasting}, J.~F., {Whitmire}, D.~P., \& {Reynolds}, R.~T. 1993, \icarus, 101, 108

\bibitem[{{Kilic} {et~al.}(2024){Kilic}, {Bergeron}, {Blouin}, {Jewett}, {Brown}, \& {Moss}}]{Kilic2024ApJ}
{Kilic}, M., {Bergeron}, P., {Blouin}, S., {et~al.} 2024, \apj, 965, 159

\bibitem[{{Kilic} {et~al.}(2006){Kilic}, {von Hippel}, {Leggett}, \& {Winget}}]{Kilic2006ApJ}
{Kilic}, M., {von Hippel}, T., {Leggett}, S.~K., \& {Winget}, D.~E. 2006, \apj, 646, 474

\bibitem[{{Kopparapu} {et~al.}(2013){Kopparapu}, {Ramirez}, {Kasting}, {Eymet}, {Robinson}, {Mahadevan}, {Terrien}, {Domagal-Goldman}, {Meadows}, \& {Deshpande}}]{Kopparapu2013ApJ}
{Kopparapu}, R.~K., {Ramirez}, R., {Kasting}, J.~F., {et~al.} 2013, \apj, 765, 131

\bibitem[{{Kozakis} {et~al.}(2020){Kozakis}, {Lin}, \& {Kaltenegger}}]{Kozakis2020ApJL}
{Kozakis}, T., {Lin}, Z., \& {Kaltenegger}, L. 2020, \apjl, 894, L6

\bibitem[{{Limbach} {et~al.}(2022){Limbach}, {Vanderburg}, {Stevenson}, {Blouin}, {Morley}, {Lustig-Yaeger}, {Soares-Furtado}, \& {Janson}}]{Limbach2022MNRAS}
{Limbach}, M.~A., {Vanderburg}, A., {Stevenson}, K.~B., {et~al.} 2022, \mnras, 517, 2622

\bibitem[{{Lin} {et~al.}(2014){Lin}, {Gonzalez Abad}, \& {Loeb}}]{Lin2014ApJLpollution}
{Lin}, H.~W., {Gonzalez Abad}, G., \& {Loeb}, A. 2014, \apjl, 792, L7

\bibitem[{{Lin} \& {Loeb}(2014)}]{Lin2014ApJLasteroid}
{Lin}, H.~W., \& {Loeb}, A. 2014, \apjl, 793, L43

\bibitem[{{Lin} {et~al.}(2022){Lin}, {Seager}, {Ranjan}, {Kozakis}, \& {Kaltenegger}}]{Lin2022ApJL}
{Lin}, Z., {Seager}, S., {Ranjan}, S., {Kozakis}, T., \& {Kaltenegger}, L. 2022, \apjl, 925, L10

\bibitem[{{Livio} {et~al.}(2005){Livio}, {Pringle}, \& {Wood}}]{Livio2005ApJL}
{Livio}, M., {Pringle}, J.~E., \& {Wood}, K. 2005, \apjl, 632, L37

\bibitem[{{Loeb} \& {Maoz}(2013)}]{Loeb2013MNRAS}
{Loeb}, A., \& {Maoz}, D. 2013, \mnras, 432, L11

\bibitem[{{Luhman} {et~al.}(2011){Luhman}, {Burgasser}, \& {Bochanski}}]{luhman2011}
{Luhman}, K.~L., {Burgasser}, A.~J., \& {Bochanski}, J.~J. 2011, \apjl, 730, L9

\bibitem[{{Manser} {et~al.}(2019){Manser}, {G{\"a}nsicke}, {Eggl}, {Hollands}, {Izquierdo}, {Koester}, {Landstreet}, {Lyra}, {Marsh}, {Meru}, {Mustill}, {Rodr{\'\i}guez-Gil}, {Toloza}, {Veras}, {Wilson}, {Burleigh}, {Davies}, {Farihi}, {Gentile Fusillo}, {de Martino}, {Parsons}, {Quirrenbach}, {Raddi}, {Reffert}, {Del Santo}, {Schreiber}, {Silvotti}, {Toonen}, {Villaver}, {Wyatt}, {Xu}, \& {Portegies Zwart}}]{Manser2019Sci}
{Manser}, C.~J., {G{\"a}nsicke}, B.~T., {Eggl}, S., {et~al.} 2019, Science, 364, 66

\bibitem[{{Mayor} {et~al.}(2011){Mayor}, {Marmier}, {Lovis}, {Udry}, {S{\'e}gransan}, {Pepe}, {Benz}, {Bertaux}, {Bouchy}, {Dumusque}, {Lo Curto}, {Mordasini}, {Queloz}, \& {Santos}}]{Mayor2011arXiv}
{Mayor}, M., {Marmier}, M., {Lovis}, C., {et~al.} 2011, arXiv e-prints, arXiv:1109.2497

\bibitem[{{Mullally} {et~al.}(2024){Mullally}, {Debes}, {Cracraft}, {Mullally}, {Poulsen}, {Albert}, {Thibault}, {Reach}, {Hermes}, {Barclay}, {Kilic}, \& {Quintana}}]{Mullally2024ApJL}
{Mullally}, S.~E., {Debes}, J., {Cracraft}, M., {et~al.} 2024, \apjl, 962, L32

\bibitem[{{Mustill} \& {Villaver}(2012)}]{Mustill2012ApJ}
{Mustill}, A.~J., \& {Villaver}, E. 2012, \apj, 761, 121

\bibitem[{{Nordhaus} \& {Spiegel}(2013)}]{Nordhaus2013MNRAS}
{Nordhaus}, J., \& {Spiegel}, D.~S. 2013, \mnras, 432, 500

\bibitem[{Oliphant(2006)}]{np}
Oliphant, T.~E. 2006, A guide to NumPy

\bibitem[{pandas~development team(2020)}]{reback2020pandas}
pandas~development team, T. 2020, pandas-dev/pandas: Pandas, doi:10.5281/zenodo.3509134

\bibitem[{{Passy} {et~al.}(2012){Passy}, {Mac Low}, \& {De Marco}}]{passy}
{Passy}, J.-C., {Mac Low}, M.-M., \& {De Marco}, O. 2012, \apjl, 759, L30

\bibitem[{{Poulsen} {et~al.}(2024){Poulsen}, {Debes}, {Cracraft}, {Mullally}, {Reach}, {Kilic}, {Mullally}, {Albert}, {Thibault}, {Hermes}, {Barclay}, \& {Quintana}}]{Poulsen2024AJ}
{Poulsen}, S., {Debes}, J., {Cracraft}, M., {et~al.} 2024, \aj, 167, 257

\bibitem[{{Rappaport} {et~al.}(2013){Rappaport}, {Sanchis-Ojeda}, {Rogers}, {Levine}, \& {Winn}}]{Rappaport2013ApJL}
{Rappaport}, S., {Sanchis-Ojeda}, R., {Rogers}, L.~A., {Levine}, A., \& {Winn}, J.~N. 2013, \apjl, 773, L15

\bibitem[{{Romero} {et~al.}(2013){Romero}, {Kepler}, {C{\'o}rsico}, {Althaus}, \& {Fraga}}]{Romero2013}
{Romero}, A.~D., {Kepler}, S.~O., {C{\'o}rsico}, A.~H., {Althaus}, L.~G., \& {Fraga}, L. 2013, \apj, 779, 58

\bibitem[{{Salaris} {et~al.}(2024){Salaris}, {Blouin}, {Cassisi}, \& {Bedin}}]{Salaris2024}
{Salaris}, M., {Blouin}, S., {Cassisi}, S., \& {Bedin}, L.~R. 2024, \aap, 686, A153

\bibitem[{{Segretain}(1996)}]{segretain1996}
{Segretain}, L. 1996, \aap, 310, 485

\bibitem[{{Shen} {et~al.}(2023){Shen}, {Blouin}, \& {Breivik}}]{Shen2023ApJL}
{Shen}, K.~J., {Blouin}, S., \& {Breivik}, K. 2023, \apjl, 955, L33

\bibitem[{{Shields} {et~al.}(2024){Shields}, {Wolf}, {Agol}, \& {Tremblay}}]{Shields2024arXiv}
{Shields}, A.~L., {Wolf}, E.~T., {Agol}, E., \& {Tremblay}, P.-E. 2024, arXiv e-prints, arXiv:2412.02694

\bibitem[{{Tremblay} {et~al.}(2011){Tremblay}, {Bergeron}, \& {Gianninas}}]{Tremblay2011}
{Tremblay}, P.~E., {Bergeron}, P., \& {Gianninas}, A. 2011, \apj, 730, 128

\bibitem[{{Tremblay} {et~al.}(2019){Tremblay}, {Fontaine}, {Gentile Fusillo}, {Dunlap}, {G{\"a}nsicke}, {Hollands}, {Hermes}, {Marsh}, {Cukanovaite}, \& {Cunningham}}]{Tremblay2019Natur}
{Tremblay}, P.-E., {Fontaine}, G., {Gentile Fusillo}, N.~P., {et~al.} 2019, \nat, 565, 202

\bibitem[{{Vanderburg} {et~al.}(2015){Vanderburg}, {Johnson}, {Rappaport}, {Bieryla}, {Irwin}, {Lewis}, {Kipping}, {Brown}, {Dufour}, {Ciardi}, {Angus}, {Schaefer}, {Latham}, {Charbonneau}, {Beichman}, {Eastman}, {McCrady}, {Wittenmyer}, \& {Wright}}]{wd1145}
{Vanderburg}, A., {Johnson}, J.~A., {Rappaport}, S., {et~al.} 2015, \nat, 526, 546

\bibitem[{{Vanderburg} {et~al.}(2020){Vanderburg}, {Rappaport}, {Xu}, {Crossfield}, {Becker}, {Gary}, {Murgas}, {Blouin}, {Kaye}, {Palle}, {Melis}, {Morris}, {Kreidberg}, {Gorjian}, {Morley}, {Mann}, {Parviainen}, {Pearce}, {Newton}, {Carrillo}, {Zuckerman}, {Nelson}, {Zeimann}, {Brown}, {Tronsgaard}, {Klein}, {Ricker}, {Vanderspek}, {Latham}, {Seager}, {Winn}, {Jenkins}, {Adams}, {Benneke}, {Berardo}, {Buchhave}, {Caldwell}, {Christiansen}, {Collins}, {Col{\'o}n}, {Daylan}, {Doty}, {Doyle}, {Dragomir}, {Dressing}, {Dufour}, {Fukui}, {Glidden}, {Guerrero}, {Guo}, {Heng}, {Henriksen}, {Huang}, {Kaltenegger}, {Kane}, {Lewis}, {Lissauer}, {Morales}, {Narita}, {Pepper}, {Rose}, {Smith}, {Stassun}, \& {Yu}}]{wd1856}
{Vanderburg}, A., {Rappaport}, S.~A., {Xu}, S., {et~al.} 2020, \nat, 585, 363

\bibitem[{{Veras}(2021)}]{Veras2021orel}
{Veras}, D. 2021, in Oxford Research Encyclopedia of Planetary Science, 1

\bibitem[{{Veras} \& {Fuller}(2020)}]{Veras2020MNRAS}
{Veras}, D., \& {Fuller}, J. 2020, \mnras, 492, 6059

\bibitem[{{W}es {M}c{K}inney(2010)}]{mckinney-proc-scipy-2010}
{W}es {M}c{K}inney. 2010, in {P}roceedings of the 9th {P}ython in {S}cience {C}onference, ed. {S}t\'efan van~der {W}alt \& {J}arrod {M}illman, 56 -- 61

\bibitem[{{Whyte} {et~al.}(2024){Whyte}, {Quiroga-Nu{\~n}ez}, {Lingam}, \& {Pinilla}}]{Whyte2024arXiv}
{Whyte}, C.~T., {Quiroga-Nu{\~n}ez}, L.~H., {Lingam}, M., \& {Pinilla}, P. 2024, arXiv e-prints, arXiv:2411.18934

\bibitem[{{Wolszczan} \& {Frail}(1992)}]{Wolszczan1992Natur}
{Wolszczan}, A., \& {Frail}, D.~A. 1992, \nat, 355, 145

\bibitem[{{Zhan} {et~al.}(2024){Zhan}, {Koll}, \& {Ding}}]{Zhan2024arXiv}
{Zhan}, R., {Koll}, D. D.~B., \& {Ding}, F. 2024, \apj, 971, 125

\bibitem[{{Zuckerman} {et~al.}(2010){Zuckerman}, {Melis}, {Klein}, {Koester}, \& {Jura}}]{Zuckerman2010ApJ}
{Zuckerman}, B., {Melis}, C., {Klein}, B., {Koester}, D., \& {Jura}, M. 2010, \apj, 722, 725

\end{thebibliography}

\end{document}